\documentclass[twocolumn,epjc3]{svjour3}  
\usepackage{axodraw2}
\smartqed  
\RequirePackage{graphicx}
\RequirePackage{amsmath,amssymb,amsfonts,slashed}
\RequirePackage[absolute]{textpos}
\RequirePackage[numbers,sort&compress]{natbib}
\RequirePackage[colorlinks,citecolor=blue,urlcolor=blue,linkcolor=blue]{hyperref}
\journalname{Eur. Phys. J. C}

\def\SOFTSUSY{{\tt SOFTSUSY}}
\def\FLEXIBLESUSY{{\tt Flex\-i\-ble\-SUSY}}
\def\FLEXIBLEEFTHiggs{{\tt Flex\-i\-ble\-EFT\-Higgs}}
\def\HIMALAYA{{\tt Himalaya}}
\def\SOFTSUSYv{\SOFTSUSY\ 4.1.1}
\def\FLEXIBLESUSYv{\FLEXIBLESUSY\ 2.1.0}
\def\HIMALAYAv{\HIMALAYA\ 1.0.1}
\def\HSSUSY{\texttt{HSSUSY}}
\def\HSSUSYv{\HSSUSY\ 2.1.0}
\def\FEYNHIGGS{\texttt{FeynHiggs}}
\def\SUSYHD{\texttt{SUSYHD}}
\newcommand{\fig}[1]{\figurename~\ref{#1}}
\newcommand{\secref}[1]{Section~\ref{#1}}
\newcommand{\SM}{\ensuremath{\text{SM}}}
\newcommand{\MSSM}{\ensuremath{\text{MSSM}}}
\newcommand{\CP}{\ensuremath{CP}}
\newcommand{\DRbar}{\ensuremath{\overline{\text{DR}}}}
\newcommand{\DRbarp}{\ensuremath{\overline{\text{DR}}'}}
\newcommand{\MSbar}{\ensuremath{\overline{\text{MS}}}}
\newcommand{\aem}{\ensuremath{\alpha_{\text{e.m.}}}}
\newcommand{\as}{\ensuremath{\alpha_{\text{s}}}}
\newcommand{\at}{\ensuremath{\alpha_t}}
\newcommand{\ab}{\ensuremath{\alpha_b}}
\newcommand{\atau}{\ensuremath{\alpha_\tau}}
\newcommand{\Qpole}{\ensuremath{Q_\text{pole}}}
\newcommand{\Qmatch}{\ensuremath{Q_\text{match}}}
\newcommand{\MS}{\ensuremath{M_S}}
\newcommand{\SQCD}{\ensuremath{\text{SQCD}}}
\newcommand{\GeV}{\ensuremath{\,\text{GeV}}}
\newcommand{\TeV}{\ensuremath{\,\text{TeV}}}
\newcommand{\DMh}{\ensuremath{\Delta M_h^{(\texttt{SS+H})}}}
\newcommand{\DMhQpole}{\ensuremath{\Delta M_h^{(\Qpole)}}}
\newcommand{\DMhQmatch}{\ensuremath{\Delta M_h^{(\Qmatch)}}}
\newcommand{\DMhMt}{\ensuremath{\Delta M_h^{(m_t)}}}
\newcommand{\DMhAlphaS}{\ensuremath{\Delta M_h^{(\as)}}}
\newcommand{\DMhAlphaEm}{\ensuremath{\Delta M_h^{(\aem)}}}
\newcommand{\DMhHSSUSY}{\ensuremath{\Delta M_h^{(\HSSUSY)}}}
\newcommand{\DMhHSSUSYytSM}{\ensuremath{\Delta M_h^{(y_t^\SM)}}}
\newcommand{\DMhHSSUSYytMSSM}{\ensuremath{\Delta M_h^{(y_t^\MSSM)}}}
\newcommand{\DMhEFT}{\ensuremath{\Delta M_h^{(v^2/\MS^2)}}}

\newcommand{\order}[1]{\ensuremath{\mathcal{O}(#1)}}

\newcommand{\QCDQED}{\ensuremath{\text{QCD}\times\text{QED}}}

\begin{document}

\begin{textblock*}{10em}(0.99\textwidth,2em)
\raggedleft\noindent\footnotesize
DAMTP-2018-04-26\\
TTK--18--14
\end{textblock*}

\title{Uncertainties in the Lightest \CP\ Even Higgs Boson Mass Prediction in the
  Minimal Supersymmetric Standard Model: Fixed Order Versus 
  Effective Field Theory Prediction
}

\author{B.C. Allanach\thanksref{addr1,e1,t1}
        \and
        A. Voigt\thanksref{addr2,e2,t2} 
}

\thankstext{e1}{e-mail: {\tt B.C.Allanach@damtp.cam.ac.uk}}
\thankstext{e2}{e-mail: {\tt alexander.voigt@physik.rwth-aachen.de}}
\thankstext{t1}{This work has been partially supported by STFC consolidated
  grant ST/P000681/1.}
\thankstext{t2}{A.~Voigt acknowledges support by the DFG Research Unit
  \textit{New Physics at the LHC} (FOR2239).}

\institute{DAMTP, CMS, University of Cambridge, Wilberforce Road, Cambridge,
  CB1 3BZ, United Kingdom \label{addr1}
           \and
           Institute for Theoretical Particle Physics and Cosmology, RWTH
           Aachen University, 52074 Aachen, Germany\label{addr2}
}


\maketitle

\begin{abstract}
We quantify and examine the uncertainties in predictions of the lightest \CP\
even Higgs boson pole mass $M_h$ in
the Minimal Supersymmetric Standard Model (\MSSM), utilising current
spectrum generators and including some three-loop corrections. There are two
broad\-ly different approximations being used: effective field theory (EFT)
where an effective Standard Model (\SM) is used below a supersymmetric mass
scale, and a fixed order calculation, where the \MSSM\ is matched to
\QCDQED\ at the electroweak scale. 
The uncertainties on the $M_h$ prediction in each approach are broken down
into logarithmic and finite pieces. The inferred values of the
stop mass parameters are
sensitively dependent upon the precision of the  prediction for $M_h$. 
The fixed order calculation appears to be more accurate below a supersymmetry (SUSY) mass
scale of $\MS \approx 1.2\ \TeV$, whereas above this scale, the EFT calculation
is more accurate. 
 We also revisit the range of the lightest stop mass across fine-tuned
 parameter space that has an appropriate stable 
 vacuum and is compatible with the lightest \CP\ even Higgs boson $h$
 being identified with the one discovered at the ATLAS and CMS experiments in
 2012; we  achieve a maximum  
 value of $\sim 10^{11}$ GeV. 
\end{abstract}

\section{Introduction}
\label{sec:intro}
The 2012 discovery at Large Hadron Collider experiments~\cite{Aad:2012tfa,Chatrchyan:2012xdj} of a resonance
that has measured 
properties compatible with those of 
a \SM\ Higgs boson, raises some expectations. Should the
language of quantum field theory be interpreted correctly by most of the
research community,
huge corrections to the Higgs boson mass are
expected, rendering its measured value \cite{Aad:2015zhl} of
\begin{equation}
M_h = 125.09\pm 0.32 \textrm{~GeV} \label{higgsMass}
\end{equation}
untenable unless its value is finely tuned with unrelated contributions
cancelling to a suspiciously large degree. This {\em technical hierarchy
  problem}~can be solved 
by new physics that appears around the TeV scale, the foremost example being
TeV scale
supersymmetry. TeV scale supersymmetry predicts that the masses of new
hitherto undiscovered sparticles are not much higher than the TeV
scale. These to date have not been discovered, and the most natural
portion of supersymmetry parameter space is being heavily squeezed by
experimental constraints. 

It is possible that there is some misunderstanding in the way that quantum
field theory generates such huge corrections and that the technical hierarchy
problem should be taken {\em cum grano salis}. It is also possible that
supersymmetry is simply a little late to the LHC party, is a little heavier
than expected and isn't quite as natural as was originally thought. It is
therefore crucial to try to discover superparticles. Within the
simplest supersymmetric extension of the \SM, the \MSSM, there is
a lot of parameter space where $h$ appears to be essentially \SM\
like. Its mass, which is $M_{h}= M_Z \cos 2\beta$ at tree level in the
decoupling limit ($M_Z$ being the mass of the $Z$ boson and
$\tan\beta=v_u/v_d$ is the ratio of the two neutral $CP-$even \MSSM\ Higgs
field vacuum expectation values (VEVs)), receives large
corrections at the loop level. It has been known for some time that the
largest corrections to its mass  (squared) come
from top/stop corrections, which are enhanced by the large value of the top
mass~\cite{Allanach:2004rh}: 
\begin{equation}
M_{h}^2 = M_Z^2 \cos^2 2 \beta + 
\frac{3}{2 \pi^2} \frac{m_t^4}{v^2} \left[ \ln \frac{M^2}{m_t^2} +
  \frac{X_t^2}{M^2} - \frac{X_t^4}{12 M^4}\right]
\label{oneLstop},
\end{equation}
where $M=\sqrt{m_{{\tilde t}_1} m_{{\tilde t}_2}}$,
$m_{{\tilde t}_i}$ is the running $i^{\text{th}}$ stop mass, $m_t$ is the running top
mass, $X_t= A_t - \mu/\tan\beta$ is the running stop mixing parameter and 
$v\sim 246\GeV$ is the running SM-like Higgs VEV. Each quantity on the right
hand side of
Eq.~\eqref{oneLstop} is evaluated
at some \DRbarp\ \cite{Siegel:1979wq,Capper:1979ns,Jack:1994rk}
renormalisation scale $Q$.  The stops 
play a crucial r\^{o}le in bringing the value of $M_h$ predicted
up to the measured value from the tree-level value.
The measured value of $M_{h}$ in Eq.~\eqref{higgsMass}
prefers those parts of parameter space that have larger stop masses and/or
large mixing between the two stops. 

The truncation of perturbation
theory at a finite order generates a
theoretical uncertainty on the prediction of $M_{h}$. This then leads to 
an associated uncertainty in the
inferred masses and mixings of stops that agree with the experimentally
inferred value of $M_{h}$.
The allowed range of stop parameter space depends very sensitively on the
accuracy of the $M_{h}$ prediction. Eq.~\eqref{oneLstop} shows that the stop
mass scale depends 
roughly exponentially upon $M_h$ in the high $m_{{\tilde t}_i}$ limit\footnote{More
  precisely, $M_{h}^2$ has a logarithmic dependence on $M$
  in the large $M$ limit.}.
Achieving the most precise prediction for $M_h$ is then of paramount
importance. 
In order to predict $M_{h}$ with higher accuracy and greater precision, higher-order
contributions and large log re-summation are required. To date, terms up to
two-loop order have been computed in the on-shell scheme~\cite{Hempfling:1993qq,Heinemeyer:1998kz,Heinemeyer:1998jw,Heinemeyer:1998np,Heinemeyer:1999be,Degrassi:2001yf,Brignole:2001jy,Dedes:2003km,Heinemeyer:2004xw,Heinemeyer:2007aq,Hollik:2014bua,Passehr:2017ufr} and
up to three-loop order in the \DRbar/\DRbarp\ scheme~\cite{Degrassi:2001yf,Brignole:2001jy,Martin:2001vx,Martin:2002iu,Martin:2002wn,Dedes:2002dy,Brignole:2002bz,Dedes:2003km,Martin:2003it,Allanach:2004rh,Martin:2004kr,Martin:2005eg,Martin:2007pg,Harlander:2008ju,Kant:2010tf,Martin:2017lqn}.

Currently, ATLAS and CMS perform many different searches for stops. They
depend upon 
various \MSSM\ parameters, but in the more constraining and direct cases,
the searches rule out lightest stop masses up to around $1\TeV$
\cite{ATLAS:2017kyf,ATLAS:2017tpg}. Ideally, it would 
be useful to determine exactly how heavy the stops might be so that it can be
judged how much of the viable parameter space has been excluded and so that
one may inform the utility of future higher energy colliders such as the
Future Circular Collider
(FCC)~\cite{Gomez-Ceballos:2013zzn,Dam:2015nir,Janot:2015mqv}. 
However, it was shown in
Refs.~\cite{Giudice:2011cg,Draper:2013oza,Bagnaschi:2014rsa,Lee:2015uza,Vega:2015fna}
that stops far 
heavier than 
the 100 TeV putative centre of mass energy of the FCC are compatible with
Eq.~\eqref{higgsMass}, provided that one is willing to accept the tuning in
$v$ implied by the technical hierarchy problem\footnote{Large stop masses make
the running soft-breaking squared Higgs mass parameters very large, requiring a huge
cancellation in the minimisation of the Higgs potential to achieve $v\sim
246~\GeV$.}. We shall repeat this calculation taking our more precise
estimates of the theoretical uncertainties in $M_h$ into account. 

\begin{figure}
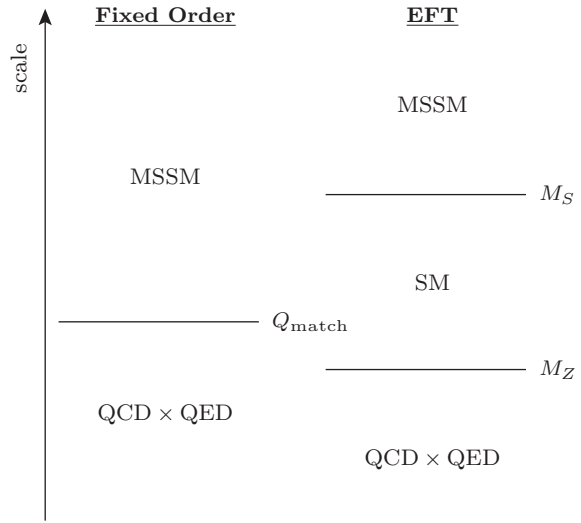

\begin{center}
\begin{axopicture}(200,200)
\Line[arrow,arrowpos=1](5,10)(5,200)
\Text(-5,190)(90)[r]{scale}
\Text(50,200)[c]{\underline{\bf Fixed Order}}
\Text(50,140)[c]{\MSSM}
\Line(10,85)(85,85)\Text(90,85)[l]{\Qmatch}
\Text(50,50)[c]{\QCDQED}
\Text(150,200)[c]{\underline{\bf EFT}}
\Text(150,167)[c]{\MSSM}
\Line(110,133)(185,133) \Text(190,133)[l]{$\MS$}
\Text(150,100)[c]{SM}
\Line(110,67)(185,67) \Text(190,67)[l]{$M_Z$}
\Text(150,33)[c]{\QCDQED}
\end{axopicture}
\end{center}
\caption{Schematics of different MSSM approximation schemes: fixed
  order \DRbarp\ and EFT. 
  In the fixed order \DRbarp\ scheme the MSSM is matched to
  effective \QCDQED\ at the scale $\Qmatch$.  In the EFT scheme the
  MSSM is matched to an effective \SM\ at the supersymmetric 
  scale $\MS$. Horizontal lines show the matching scales.}
  \label{fig:eftvFO}
\end{figure}

There are several current publicly available MSSM spectrum calculator computer
programs on the market. These calculate the spectrum consistent with
weak-scale data on the gauge couplings and the masses of SM states. Each
employs an approximation scheme. 
The two approximation schemes examined here are called the fixed order
\DRbarp\ scheme
and the EFT scheme, depicted in Fig.~\ref{fig:eftvFO}. The fixed order \DRbarp\ scheme
matches effective QED$\times$QCD to the \MSSM\ at a single scale \Qmatch.
The values $\Qmatch=M_Z$
or $\Qmatch=M_t$ are commonly taken, and data on gauge couplings and quark,
lepton and electroweak boson masses are input at this scale \Qmatch\ (see
Ref.~\cite{Allanach:2001kg} for a more detailed description). These couplings
are then run using \MSSM\ renormalisation group equations (RGEs) to $M$,
where the Higgs potential minimisation 
conditions are imposed and supersymmetric physical observables including $M_h$
are calculated. $M_h$ is calculated using the known higher order diagrammatic
corrections, up to three loops, of the order
$o \in\{\as\at$, $\as\ab$, $\at^2$, $\at\ab$, $\ab^2$, 
$\atau^2$, $\as^2\at\}$, where $\as = g_3^2/(4\pi)$,
$\alpha_{t,b,\tau} = y_{t,b,\tau}^2/(4\pi)$ and $y_t$, $y_b$, $y_\tau$
are the top, bottom and tau Yukawa couplings, respectively, and $g_3$
is the QCD gauge coupling. These fixed-order corrections include
two-loop terms which are proportional to $o \ln^2 (\MS/M_Z)/(4 \pi)^2$ as 
well as terms of order $o M_Z^2 / \MS^2 / (4 \pi)^2$.
However, some three-loop terms, for example of order
$\{\at^2\as, \at^3\}$ $\times$ $\ln^3 (\MS/M_Z) / (4 \pi)^3$, are missed.
As $\MS$ becomes larger (for example as motivated by the negative results of
sparticle searches), 
such missing logarithmic higher order terms become numerically more important, and 
missing them will imply a larger uncertainty in the fixed order \DRbarp\ prediction of
$M_h$. 
This has motivated the approximation scheme which we call the EFT
scheme, where the heavy SUSY particles are decoupled at the SUSY scale
$\MS$ and the RGEs are used to re-sum the large logarithmic
corrections. However, the EFT scheme neglects terms of order
$M_Z^2/\MS^2$ at the tree level and therefore is less
accurate the closer $\MS$ is to $M_Z$. Which scheme is the most
accurate for various different physical predictions is not obvious
beforehand and depends on the MSSM parameters. It is one
of our goals to determine in which domain of $\MS$ the fixed-order
scheme becomes less accurate than the EFT scheme.

The preceding paragraph has been greatly simplified for clarity of
discussion. In the \MSSM\ there are many gauge and Yukawa couplings and
one-loop corrections from all of these are included in the fixed order \DRbarp\
calculations. Also, we have used $\MS$ as a catch-all supersymmetric scale,
but really the individual sparticles contribute to the logarithms and finite
terms with their own masses, not with some universal value of $\MS$.

The programs used for our $M_{h}$ predictions are the fixed order
\DRbarp\ spectrum generators 
\SOFTSUSYv{}~\cite{Allanach:2001kg,Allanach:2014nba}, 
\FLEXIBLESUSYv{}~\cite{Athron:2014yba,Athron:2017fvs} 
and \HSSUSYv\ \cite{Athron:2017fvs}, which uses the EFT approach.
We include the three-loop corrections that are 
available in \HIMALAYAv{}~\cite{Harlander:2017kuc}. 

In 
Ref.~\cite{Bahl:2017aev}, the hybrid fixed order \DRbarp\/EFT calculation of \FEYNHIGGS\
\cite{Hahn:2013ria,Bahl:2016brp}
was compared to the purely EFT calculation of \SUSYHD\ \cite{Vega:2015fna}.
The observed numerical differences between the
(mostly) on-shell hybrid calculation of \FEYNHIGGS\ and the \DRbarp\
calculation of \SUSYHD\ were found to be mainly caused by
renormalisation scheme conversion terms, the treatment of higher-order
terms in the determination of the Higgs boson pole mass and the
parametrisation of the top Yukawa coupling.
When these differences are taken into consideration,
excellent agreement was found between the two programs for SUSY scales
above $1\TeV$.  This finding confirms that above this scale the terms
neglected in the EFT calculation are in fact negligible and the EFT
calculation yields an accurate prediction of the Higgs boson mass.
Similarly, in Ref.~\cite{Athron:2016fuq} the \DRbarp\ hybrid fixed
order/EFT calculation implemented in \FLEXIBLESUSY\ (denoted as
\FLEXIBLEEFTHiggs) was compared to the \DRbarp\ fixed order
calculation available in \FLEXIBLESUSY.  A prescription for an
uncertainty estimation of both calculations was given and it was found
that (based on that uncertainty estimate) above a few TeV the hybrid
and the pure EFT calculations are more precise than the fixed order \DRbarp\
calculation.

Our work differs from Refs.~\cite{Athron:2016fuq,Bahl:2017aev} in that
we perform a comparison between the \DRbarp\ fixed order and the pure
EFT predictions. Our \DRbarp\ fixed order calculation is also a loop higher in
order than the previous work.
We shall give a prescription for the
estimation of the theoretical uncertainties of the two $M_h$
predictions in the \DRbarp\ scheme based on the procedures described
in Ref.~\cite{Bagnaschi:2014rsa,Vega:2015fna,Athron:2016fuq}.  Based
on our uncertainty estimates we derive an $\MS$ region in that scheme,
above which the EFT prediction becomes more precise than the fixed
order one.

In section~\ref{sec:1}, we estimate and dissect theoretical uncertainties in
state-of-the 
art predictions of the lightest
\CP\ even Higgs boson pole mass in the \DRbarp\ scheme.
Then, in section~\ref{sec:2}, we update the 
upper bounds on the lightest stop mass from the experimental determination of
the Higgs boson mass and  
from the stability of an appropriate vacuum by our detailed quantification of
the theoretical uncertainties and state-of-the-art calculation of $M_h$. We
summarise in section~\ref{sec:3}.

\section{Higgs boson Mass Prediction Uncertainties}
\label{sec:1}

Sources of uncertainty in the \DRbarp\ fixed-order
calculation of the lightest \CP-even Higgs boson pole mass prediction
can be divided into two classes:
\begin{itemize}
\item Missing higher order contributions to the Higgs self energy and
  to the electroweak symmetry breaking (EWSB) conditions.
\item Missing higher order corrections in the determination of the
  running  \DRbarp\ gauge and Yukawa couplings and the 
  VEVs from experimental quantities.
\end{itemize}
The prescription presented in Ref.~\cite{Athron:2016fuq} to estimate
these missing higher order contributions is sensitive to leading and
subleading logarithmic as well as non-loga\-rith\-mic terms.  An
analysis of how these different kinds of higher order terms enter the
uncertainty estimate can be found in that reference.
In the \CP-conserving \MSSM\ the known two- and three-loop
contributions to the \CP\ even Higgs self energy and EWSB conditions
are included.  The currently unknown (subleading) logarithmic higher order
corrections can be estimated by varying the renormalisation scale at which the
Higgs boson mass is calculated, \Qpole.  We estimate this
uncertainty as in Ref.~\cite{Athron:2016fuq},
\begin{align}
  \DMhQpole = \max_{\Qpole\in[\MS/2,2\MS]}\left|M_h(\Qpole) - M_h(\MS)\right| \,,
\end{align}
where \MS\ is the SUSY scale, usually set to
$\MS = \sqrt{m_{\tilde{t}_1}m_{\tilde{t}_2}}$.
\begin{figure}
  \includegraphics[width=\columnwidth]{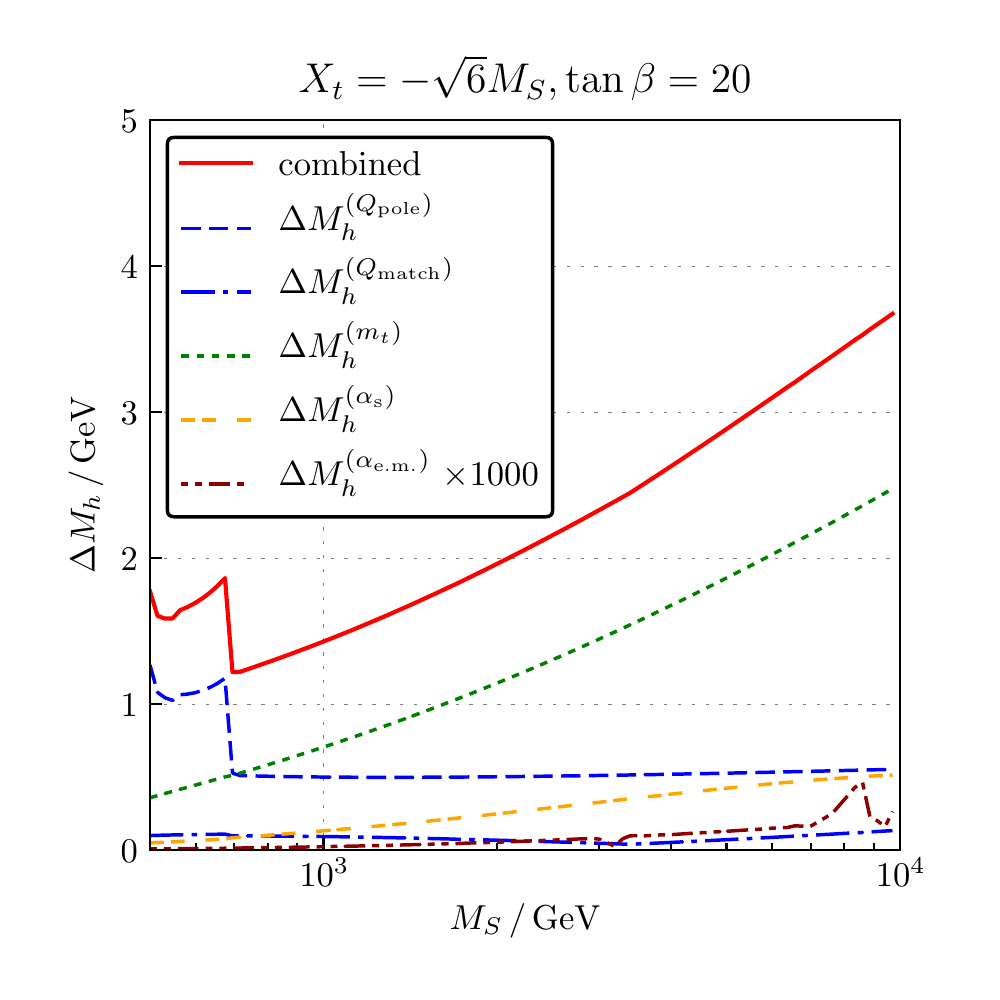}
  \caption{Individual sources of uncertainty of the three-loop fixed order \DRbarp\
    Higgs boson mass prediction of \SOFTSUSY.}
  \label{fig:2}
\end{figure}
In \fig{fig:2} we show this uncertainty as the blue dashed line for a
scenario with degenerate \DRbarp\ mass parameters (aside from the Higgs mass
parameters, which are fixed in order to achieve successful EWSB), $\tan\beta = 20$ and
maximal stop mixing, $X_t = -\sqrt{6}\MS$.  For this scenario
$\DMhQpole$ varies between $0.5$--$1\GeV$, depending on the SUSY
scale.  In Ref.~\cite{Athron:2016fuq} this uncertainty is larger,
because the three-loop contribution to the Higgs boson mass was not included. 
The kink at $\MS \approx 700\GeV$ is due to a switch in the
approximation scheme being used in the calculation of the three-loop
contribution of \HIMALAYA: the integrands of the three-loop integrals
were expanded in different sparticle ``mass hierarchies'' where different
sparticles were approximated as being massless~\cite{Kant:2010tf}.  As $\MS$
changes, \HIMALAYA\ switches from 
one mass hierarchy to another one that fits better to the given parameter
point, resulting in the kink.
We note that $\Delta M_h^{(Q_\text{pole})}$ is approximately independent of 
$M_S$ as $M_S$ becomes large. The dominant $Q_\text{pole}$ dependence comes from
the first term on the right-hand side of Eq.~\eqref{oneLstop}: 
from the running $Z$ mass, at one-loop order. This will be cancelled to
leading order in $\log(Q_\text{pole})$ by the one-loop electroweak corrections
that are added to $M_h$ by the spectrum generators that we employ. However,
higher order logarithms (formally at the two-loop order) in the electroweak
gauge couplings remain. 
These remaining pieces have no
explicit dependence at 
leading order on $M=M_S$. 
In the limit of large $M_S$, the first term in the square brackets of
Eq.~\eqref{oneLstop} contains both a dependence on a large $M_S$ and
$Q_\text{pole}$ through renormalisation of the running top Yukawa coupling
$y_t = \sqrt{2} m_t / (v \sin \beta)$. The $Q_\text{pole}$ dependence of leading logarithm terms due 
to this are
cancelled by the explicit two-loop terms of order $\alpha_t^2/(4 \pi)^2$ in
the $M_h$ calculation that the spectrum
generators employ, but higher powers of the logarithms do not cancel.
The $Q_\text{pole}$ dependence from
this term is then
formally of three-loop order, but is boosted somewhat by the large value of
$y_t$. For $\tan \beta=20$ and large $M_S$, the $Q_\text{pole}$ dependence is 
small, partly aided by cancellations in the beta function of $y_t$. However, for $\tan \beta=5$, as is the
case in Ref.~\cite{Athron:2016fuq}, for example, one can see an increase in 
scale uncertainty with a larger $M_S$ due to a larger value of $y_t$
(and consequently a larger beta function/scale dependence).

The size of the missing (subleading) logarithmic high\-er order contributions to the
running \MSSM\ \DRbarp\ gauge and Yukawa couplings can be estimated in a
similar way to that of $Q_\text{pole}$ by varying the renormalisation scale
\Qmatch, at which the
said parameters are determined.  We define this uncertainty as
\begin{align}
  \DMhQmatch = \max_{Q\in[\Qmatch/2,2\Qmatch]}\left|M_h(Q) - M_h(\Qmatch)\right|
\end{align}
where $\Qmatch$ is the scale at which these parameters are
determined, usually set to $M_Z$ or $M_t$.  The uncertainty $\DMhQmatch$ is
shown as blue dashed-dotted line in \fig{fig:2} and is below $0.2\GeV$
for the scenario shown.

Besides the logarithmic higher order corrections there are also
`non-logarithmic' higher corrections, which are important and should be
taken into account in any robust uncertainty estimate
\cite{Athron:2016fuq}.  We estimate these non-log\-a\-rith\-mic corrections
by changing the calculation of the running \MSSM\ parameters $m_t$,
$\as$ and $\aem$ by higher orders: the running \DRbarp\ top mass $m_t$
is calculated in two different ways, similar to Ref.~\cite{Athron:2016fuq}: 
\begin{align}
m_t^{(1)} &=
M_t + \widetilde{\Sigma}_t^{(1),S} +
M_t \left[
  \widetilde{\Sigma}_t^{(1),L} +
  \widetilde{\Sigma}_t^{(1),R}
\right] \nonumber \\
&\phantom{={}} + M_t
\left[\widetilde{\Sigma}_t^{(1),\SQCD}
+ \widetilde{\Sigma}_t^{(2),\SQCD}
+ \left(\widetilde{\Sigma}_t^{(1),\SQCD}\right)^2
\right]
\label{eq:mt_SS}
\end{align}
and
\begin{align}
m_t^{(2)} &=
M_t + \widetilde{\Sigma}_t^{(1),S} +
m_t \left[
  \widetilde{\Sigma}_t^{(1),L} +
  \widetilde{\Sigma}_t^{(1),R}
\right] \nonumber \\
&\phantom{={}} +
m_t
\left[\widetilde{\Sigma}_t^{(1),\SQCD} +
\widetilde{\Sigma}_t^{(2),\SQCD}
\right] \,,
\label{eq:mt_SP}
\end{align}
where $M_t$ denotes the top pole mass, $\widetilde{\Sigma}_t^{(1),S}$,
$\widetilde{\Sigma}_t^{(1),L}$ and $\widetilde{\Sigma}_t^{(1),R}$
denote the scalar, left-handed and right-handed part of the one-loop top
self energy without SUSY-QCD contributions and
$\widetilde{\Sigma}_t^{(1,2),\SQCD}$ denote the one-loop and two-loop SUSY-QCD
contributions from
Refs.~\cite{Avdeev:1997sz,Bednyakov:2002sf,Bednyakov:2005kt}\footnote{Note
  that the terms 
involving square brackets differ in
Eqs.~\eqref{eq:mt_SS},\eqref{eq:mt_SP}.}.
Note, that in Ref.~\cite{Athron:2016fuq} the two-loop SM-QCD
contribution has been used on the right hand side of
Eqs.~\eqref{eq:mt_SS}--\eqref{eq:mt_SP}, while here we use the full
two-loop SUSY-QCD contribution of $\order{\alpha_s^2}$.  Since the
latter is significantly larger, the difference between $m_t^{(1)}$ and
$m_t^{(2)}$ is larger in our case.
Eqs.~\eqref{eq:mt_SS} and \eqref{eq:mt_SP} are equivalent at
$\order{\as^2}$, but differ at $\order{\aem\as}$, for example.
Since the difference contains both logarithmic and non-logarithmic
terms, it can be used as an uncertainty estimate.  Similar to
Ref.~\cite{Athron:2016fuq} we define
\begin{align}
  \DMhMt = \left|M_h(m_t^{(1)})
    - M_h(m_t^{(2)})\right| \,.
\end{align}
In Ref.~\cite{Athron:2016fuq} four different top mass
calculations are combined, whilst we combine only two.
The size of $\DMhMt$ is shown in \fig{fig:2} as a green dotted line.
Since $\DMhMt$ contains terms of the form $\log(m_t/m_{\tilde{t}_i})$,
the uncertainty increases logarithmically with the SUSY scale.  It
therefore serves as an estimate of both (leading) logarithmic and
non-logarithmic higher order corrections and is a reasonable measure
to express the fact that the fixed-order calculation suffers from a
large theoretical uncertainty for multi-TeV stop masses.

We estimate the
effect of unknown higher order 
logarithmic and non-logarithmic threshold corrections to $\as$ and
$\aem$ in a similar way to our approach for estimating $\Delta M_h^{(m_t)}$:
\begin{align}
  \as^{(1)}  &= \frac{\as^{\SM(5)}}{1 - \Delta^{(1)}\as - \Delta^{(2)}\as} \,, \label{eq:as_SS}\\
  \as^{(2)}  &= \as^{\SM(5)} \left[1 + \Delta^{(1)}\as + (\Delta^{(1)}\as)^2 + \Delta^{(2)}\as\right] \,, \label{eq:as_SS_alt}
\end{align}
and
\begin{align}
  \aem^{(1)} &= \frac{\aem^{\SM(5)}}{1 - \Delta^{(1)}\aem -
         \Delta^{(2)}\aem} \,, \label{eq:aem_SS} \\
  \aem^{(2)} &= \aem^{\SM(5)} \Big[1 + \Delta^{(1)}\aem + (\Delta^{(1)}\aem)^2 \nonumber \\
  &\qquad\qquad\quad + \Delta^{(2)}\aem\Big] \label{eq:aem_SS_alt}
\end{align}
and take the difference as an uncertainty estimate,
\begin{align}
  \DMhAlphaS &= \left| M_h(\as^{(1)}) - M_h(\as^{(2)}) \right| \,, \\
  \DMhAlphaEm &= \left| M_h(\aem^{(1)}) - M_h(\aem^{(2)}) \right| \,.
\end{align}
Note that the uncertainties estimated by $\DMhAlphaS$ and
$\DMhAlphaEm$ were not included in
Ref.~\cite{Athron:2016fuq}.  Their respective sizes are shown in \fig{fig:2}
as yellow dashed and brown double-dotted lines, respectively.  Due to
the logarithmic contributions of the form $\log(m_t/m_{\tilde{t}_i})$
to the threshold corrections of \as\ and \aem, the two uncertainties
increase logarithmically with the SUSY scale.  However, since \aem\ is
very small, the uncertainty $\DMhAlphaEm$ is negligible.
The magnitude of $\DMhAlphaS$ can be around $20\%$ of $\DMhMt$
for large $M_S$.

There are some inter-dependencies between the different sources of uncertainty
and it is practically impossible to exactly take these into account unless the
higher order corrections are explicitly calculated. However, the quantification of
theoretical uncertainties is an inexact pursuit and it serves us well enough
to combine the different sources of uncertainty linearly
\begin{align}
  \DMh &= \DMhQpole + \DMhQmatch + \DMhMt \nonumber \\
  &\quad + \DMhAlphaS + \DMhAlphaEm \,
\end{align}
in order to have some kind of reasonable estimate of the total level of
theoretical uncertainty in the prediction.
The combination \DMh\ is shown in \fig{fig:2} as a red solid line.  As
expected, due to logarithmic contributions of the form
$\log(m_t/m_{\tilde{t}_i})$, the combined uncertainty of the
fixed-order calculation of \SOFTSUSY\ increases
with the SUSY scale and reaches $\DMh \sim 4\GeV$ for
$\MS \sim 10\TeV$.
The size of the individual uncertainties show that the prescription
proposed in Ref.~\cite{Athron:2016fuq} is reasonable, because the
additional uncertainties $\DMhQmatch$, $\DMhAlphaS$ and $\DMhAlphaEm$
that we have introduced here are small.  However, compared to the
combined uncertainty estimate of Ref.~\cite{Athron:2016fuq} our
combined uncertainty is smaller by about $40\%$ for SUSY scales of
around $\MS \sim 1\TeV$ and about $25\%$ smaller for
$\MS \sim 10\TeV$.  The main reasons are the reduced scale uncertainty
$\DMhQpole$ due to the three-loop Higgs boson mass
corrections that are included here and our different definition of $\DMhMt$. 

In the following we compare the fixed-order Higgs boson mass prediction
for this scenario to the pure EFT calculation
of \HSSUSY\ \cite{Athron:2017fvs}.  \HSSUSY\ is a spectrum generator
from the \FLEXIBLESUSY\ package, which implements the high-scale SUSY
scenario, where the quartic \SM\ Higgs coupling
$\lambda(\MS)$ is predicted from matching to the \MSSM\ at a high SUSY
scale $\MS$.  It provides a prediction of the Higgs pole mass in the
\MSSM\ in the pure EFT limit, $v^2 \ll \MS^2$, up to the two-loop level
$\order{\as(\at+\ab) + (\at+\ab)^2 + \atau\ab + \atau^2}$
\cite{Degrassi:2012ry,Martin:2014cxa,Bagnaschi:2014rsa,Bagnaschi:2017xid},
including next-to-next-to-leading-log (NNLL) re-summation
\cite{Bednyakov:2013eba,Buttazzo:2013uya}.  Additional pure \SM\
three- and four-loop corrections
\cite{Chetyrkin:1999qi,Melnikov:2000qh,Chetyrkin:2000yt,Martin:2015eia,Chetyrkin:2016ruf,Bednyakov:2015ooa}
can be taken into account on demand.

\begin{figure}
  \includegraphics[width=\columnwidth]{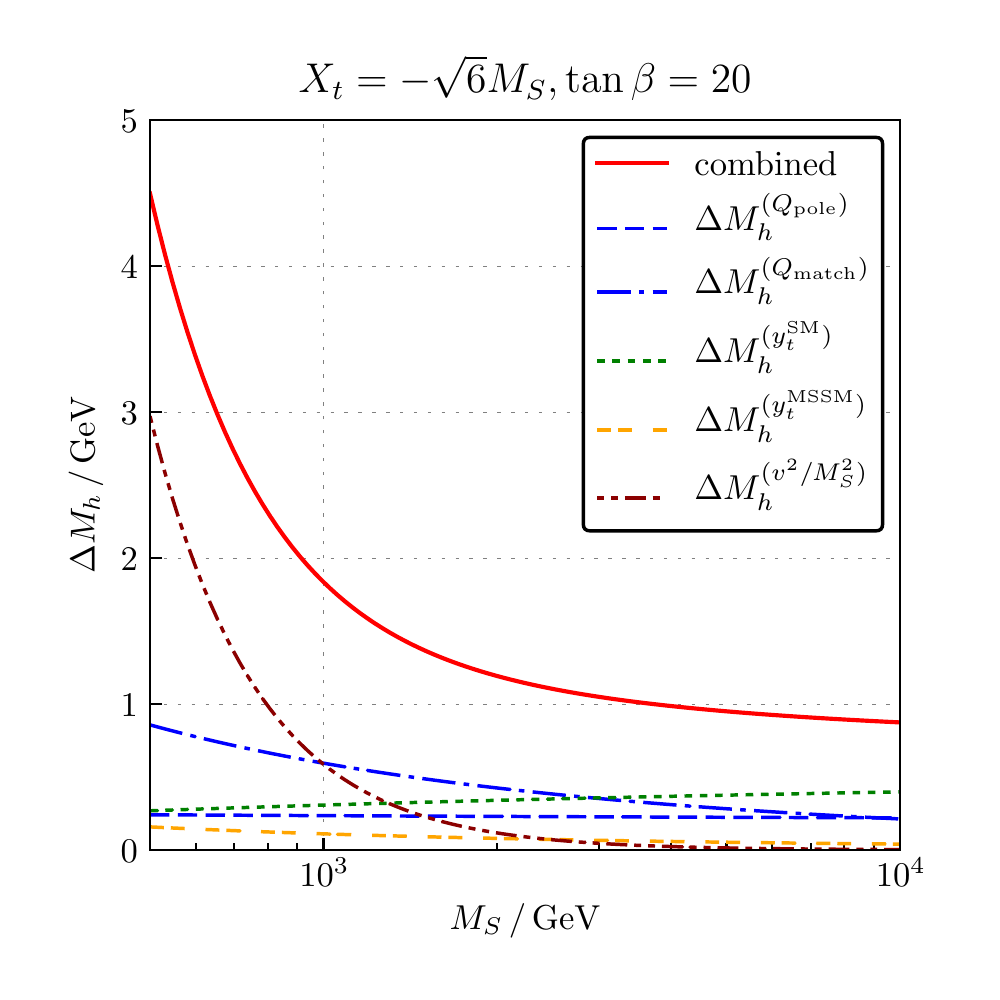}
  \caption{Individual sources of uncertainty of the two-loop EFT Higgs
    boson mass prediction of \HSSUSY.}
  \label{fig:DMh_HSSUSY}
\end{figure}

To estimate the Higgs boson mass uncertainty of \HSSUSY\ we use the
procedure developed in Ref.~\cite{BVW}, which is an extension of the
methods used in Refs.~\cite{Bagnaschi:2014rsa,Vega:2015fna}.  The sources of uncertainty
of \HSSUSY\ are divided into the following three categories:
\begin{itemize}
\item \emph{\SM\ uncertainty} from missing higher order
  corrections in the determination of the running \SM\
  \MSbar\ parameters
\item \emph{EFT uncertainty} from neglecting terms of order\\
  $\order{v^2/\MS^2}$
\item \emph{SUSY uncertainty} from missing higher order contributions
  from SUSY particles
\end{itemize}
As in the fixed order \DRbarp\ calculation, we divide the \emph{\SM\
  uncertainty} into a logarithmic and non-logarithmic part.
However, since large logarithmic corrections to the Higgs mass are
re-summed in the EFT calculation, for the `logarithmic part',
we refer specifically to smaller logarithms of the form
$\ln(\Qmatch/m_{\tilde{t}_1})$ or $\ln(\Qpole/M_t)$.  These small
logarithmic higher order corrections are estimated by varying the
renormalisation scale $\Qpole$, at which the Higgs boson mass is
calculated in the effective \SM:
\begin{align}
  \DMhQpole = \max_{\Qpole\in[M_t/2,2M_t]}\left|M_h(\Qpole) - M_h(M_t)\right| \,.
\end{align}
The non-logarithmic part is estimated by switching the
three-loop QCD contributions \cite{Chetyrkin:1999qi,Melnikov:2000qh} on or off
in the extraction of the running \SM\ top Yukawa coupling
from the top pole mass,
\begin{align}
  \DMhHSSUSYytSM = \left| M_h(y_t^{\SM,2\ell}(M_Z)) - M_h(y_t^{\SM,3\ell}(M_Z)) \right| \,.
\end{align}
Although this difference is sensitive to non-logarithmic higher order
contributions to the Higgs boson mass, it shows an additional
dependence on the separation of the electroweak scale and the SUSY scale (as
was observed in Refs.~\cite{Vega:2015fna,Braathen:2017jvs} and 
shown in the green dotted line of Fig.~\ref{fig:DMh_HSSUSY}).
The main reason for this dependence is that the running top Yukawa coupling
(the largest dimensionless parameter in the MSSM) enters
the RGEs of the other MSSM parameters, thus affecting their running
below $\MS$.  The effect is stronger for more separated scales. 

The \emph{EFT uncertainty} is estimated by Ref.\
\cite{Vega:2015fna} by multiplying the one-loop
contribution of each individual SUSY particle to the quartic Higgs
coupling $\lambda(\MS)$ at the SUSY scale by the factor
$(1 + v^2/\MS^2)$.  We use the resulting change in the Higgs boson
pole mass prediction as an estimate for the \emph{EFT uncertainty},
\begin{align}
  \DMhEFT = \left| M_h - M_h(v^2/\MS^2) \right| \,,
\end{align}
where $M_h(v^2/\MS^2)$ is the predicted Higgs boson mass with the
additional $v^2/\MS^2$ terms.  In
Ref.~\cite{Bagnaschi:2017xid} it was shown that this uncertainty
estimate is very conservative.

The \emph{SUSY uncertainty} is also
divided into a logarithmic and a `non-logarithmic' part.  We estimate
the (leading) logarithmic part again 
by varying the scale \Qmatch, at which the
matching of the MSSM to the effective \SM\ is performed,
similar to Ref.~\cite{Vega:2015fna},
\begin{align}
  \DMhQmatch = \max_{Q\in[\MS/2,2\MS]}\left|M_h(Q) - M_h(\MS)\right| \,.
\end{align}
Like \DMhHSSUSYytSM, \DMhQmatch\ also shows an additional
dependence on the separation of the electroweak scale and the SUSY scale 
due to the dependence of the RGEs on the running parameters.
The non-logarithmic part is estimated by re-pa\-ra\-me\-tri\-sing the
threshold correction for $\lambda(\MS)$ in terms of the MSSM top
Yukawa coupling at the SUSY scale, $y_t^\MSSM(\MS)$, and we take the
resulting shift in the Higgs boson mass as an estimate for the
uncertainty
\begin{align}
  \DMhHSSUSYytMSSM = \left| M_h - M_h(y_t^\MSSM(\MS)) \right| \,.
\end{align}
A similar uncertainty was defined in Ref.~\cite{Athron:2016fuq}, where
the loop order of the calculation of $y_t^\MSSM(\MS)$ was switched
between tree- and one-loop level.  Our uncertainty $\DMhHSSUSYytMSSM$
is significantly smaller than the one used in
Ref.~\cite{Athron:2016fuq}, because 
we are working at one loop higher order and 
the uncertainty there contains large two-loop
next-to-leading logarithms (see the discussion in
Ref.~\cite{Athron:2017fvs}).

Analogously to our procedure with the fixed order \DRbarp\ calculation, we combine all
individual \HSSUSY\ uncertainties linearly,
\begin{align}
  \DMhHSSUSY &= \DMhQpole + \DMhQmatch + \DMhHSSUSYytSM \nonumber \\
  &\quad + \DMhHSSUSYytMSSM + \DMhEFT \,.
\end{align}
Fig.~\ref{fig:DMh_HSSUSY} shows the individual uncertainties of
\HSSUSY\ from these three categories.
For low SUSY scales, $\MS \lesssim 1\TeV$, the combined uncertainty estimate of
\HSSUSY, \DMhHSSUSY, (red solid line) is dominated by the \emph{EFT uncertainty} \DMhEFT\
(brown dashed-double-dotted line) due to the fact that the neglected
terms of $\order{v^2/\MS^2}$ are not negligible in this region.  For
$\MS \gtrsim 2\TeV$ the \emph{EFT uncertainty} drops below $0.1\GeV$
and the remaining sources dominate.  For even higher scales of
$\MS\gtrsim 10\TeV$, the two components of the \emph{SUSY
  uncertainty}, \DMhQmatch\ and \DMhHSSUSYytMSSM, become smaller
because the dimensionless running couplings become smaller at higher
SUSY scales in this scenario.  For high scales of $\MS\gtrsim 10\TeV$
the combined uncertainty is dominated by the \emph{\SM\
  uncertainty}, in particular by the uncertainty \DMhHSSUSYytSM\ in the
extraction of 
the running \SM\ top Yukawa coupling at the electroweak
scale, which remains at $\DMhHSSUSYytSM \sim 0.5\GeV$.

\begin{figure}
  \includegraphics[width=\columnwidth]{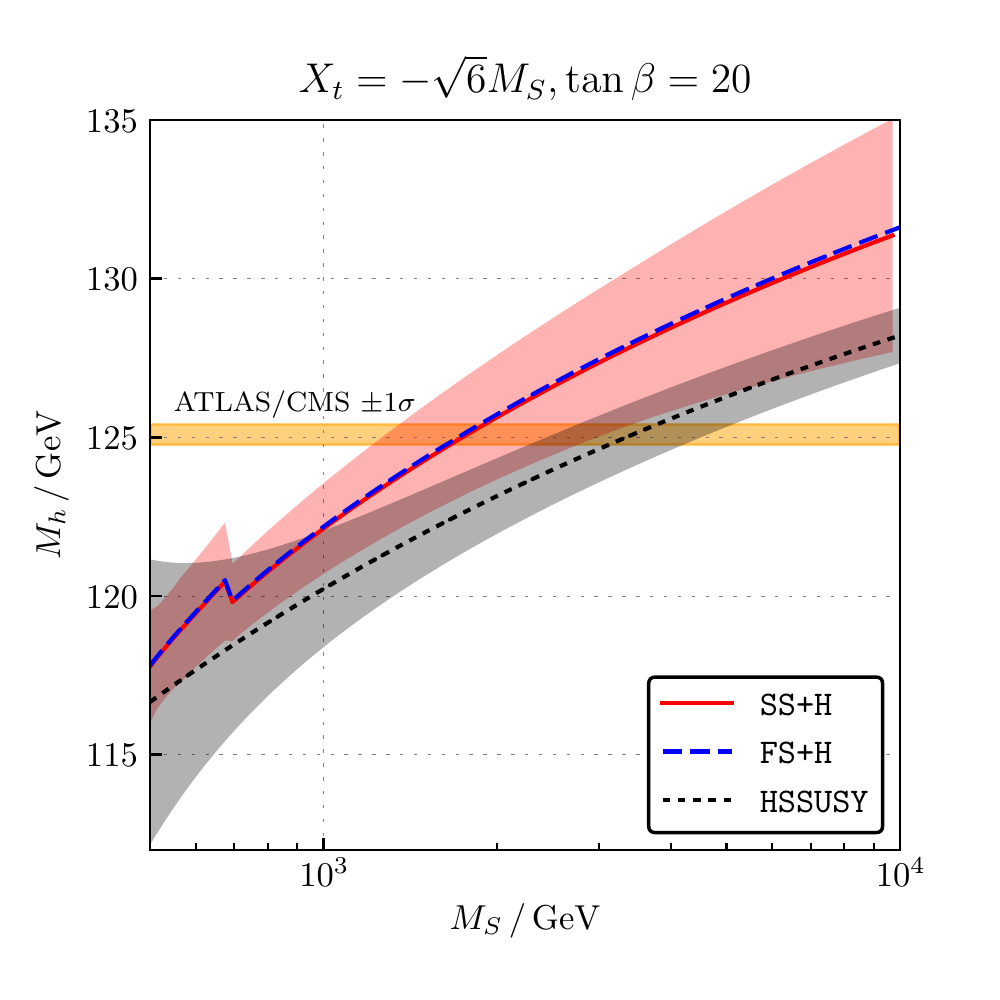}
  \caption{Higgs boson mass predictions at fixed three-loop order with
    \SOFTSUSY\ (red solid line) and \FLEXIBLESUSY\ (blue dashed line)
    and in the EFT (black dashed line). The coloured regions show the
    estimated composite theoretical uncertainties in the different
    predictions of $M_h$. The \FLEXIBLESUSY\ uncertainty is very
    similar to the \SOFTSUSY\ one, and so is not shown for reasons of
    clarity.  The orange band shows the experimentally measured
    Higgs boson mass with the experimental uncertainty.}
  \label{fig:1}
\end{figure}
In \fig{fig:1} the $M_h$ prediction in the fixed-order and
the EFT approximation schemes are shown, together with their uncertainties%
\footnote{There are small differences in the calculations of
  \SOFTSUSY\ and of \FLEXIBLESUSY\ producing their $M_h$ predictions:
  for example, the two-loop calculations of the electroweak
  corrections differ.}.
We see from the figure that the allowed $\MS$ range depends sensitively on the
approximation scheme: $1.3$--$3.0\TeV$ for fixed-order and $2.5$--$4.6\TeV$ for EFT.
The Higgs mass increases as a function of the SUSY scale due to the
logarithmic enhancement from heavy SUSY particles.  As discussed above, the
combined uncertainty of the fixed-order calculations (red band) tends
to increase with increasing $\MS$, while the uncertainty of the EFT
calculation (grey band) 
decreases.  The point where the fixed-order and the EFT calculation
have the same estimated uncertainty is $\MS^\text{equal}=1.2\TeV$.  To improve the 
prediction near this point, a ``hybrid'' calculation should
be used, where the large logarithms are re-summed and
$\order{v^2/\MS^2}$ terms are included
\cite{Athron:2017fvs,Bahl:2016brp,Bahl:2017aev,Athron:2016fuq,Bagnaschi:2017xid,Staub:2017jnp,Bahl:2018jom}.

\section{Upper bound on the lightest stop mass}
\label{sec:2}
The logarithmic enhancement of the loop corrections to the Higgs boson mass
from heavy stops suggests that there is an upper limit on the mass of
the lightest stop from the requirement of predicting
$M_h = 125.09\GeV$ and a stable and appropriate (i.e.\ colour and charge
preserving) vacuum. As was already shown in
Ref.~\cite{Bagnaschi:2014rsa}, the maximum lightest stop mass is around
$10^{11}$ GeV. 
At very large stop masses, EWSB is fine-tuned, despite the fact that the Higgs
mass in our spectrum generators comes out to be small. This is because the
generators implicitly tune parameters in order to obtain the measured central
value of the $Z$ boson mass $M_Z=91.1876$ GeV (or equivalently, the inferred
value of $v \sim 246\GeV$). We see this in the MSSM EWSB
equation~\cite{Pierce:1996zz} which predicts $M_Z$:
\begin{equation}
\frac{M_{Z}^2}{2}=
\frac{m_{\bar{H}_1}^2(Q) -  m_{\bar{H}_2}^2(Q) \tan^2
  \beta(Q)}{\tan^2 \beta(Q) - 1} 
- \frac{1}{2}  \Re\Pi_{ZZ}^T(Q) - \mu^2
\label{mucond} 
\end{equation}
where $m_{\bar{H}_i}^2 = m_{H_i}^2 - t_i/v_i$, $
\Re\Pi_{ZZ}^T(Q)$ is the $Z$ self-energy and
$t_i$ are the tadpole contributions from loops. $Q$ is the scale at which EWSB
is calculated: usually around the TeV scale and
$v_i$ are the two Higgs VEVs
of the \CP\ even electrically neutral \MSSM\ Higgs fields. 
When the stop masses are huge, they contribute to huge values of
$m_{\bar{H}_i}^2$ 
through the MSSM RGEs, which at one loop order
are~\cite{Martin:1993zk}: 
\begingroup
\allowdisplaybreaks
\begin{eqnarray}
\frac{1}{\kappa} \frac{d m_{H_1}^2}{d t} &=& 6\, y_b^2 \left(m_{H_1}^2 + m_{{\tilde
                                         Q}_3}^2 + m_{{\tilde d}_3}^2
                                         +A_b^2\right)
                                         \nonumber \\ &&
                                         - 6 g_2^2 M_2^2
  - \frac{6}{5}g_1^2M_1^2 + \frac{3}{5}g_1^2\Big[m_{H_2}^2 - m_{H_1}^2\nonumber\\ &&
                                         +\text{Tr}(m_{\tilde Q}^2 - m_{\tilde
                                         L}^2 - 2 m_{\tilde u}^2 + m_{\tilde
                                         d}^2 + m_{\tilde e}^2)\Big], \\
\frac{1}{\kappa} \frac{d m_{H_2}^2}{d t} &=& 6\, y_t^2 \left(m_{H_2}^2 + m_{{\tilde
                                         Q}_3}^2 + m_{{\tilde u}_3}^2
                                         +A_t^2\right)
                                         \nonumber \\ &&
                                         - 6 g_2^2 M_2^2
  - \frac{6}{5}g_1^2M_1^2 + \frac{3}{5}g_1^2\Big[m_{H_2}^2 - m_{H_1}^2\nonumber\\ &&
                                         +\text{Tr}(m_{\tilde Q}^2 - m_{\tilde
                                         L}^2 - 2 m_{\tilde u}^2 + m_{\tilde
                                         d}^2 + m_{\tilde e}^2)\Big],
\end{eqnarray}%
\endgroup
where $t$ is the natural logarithm of
the renormalisation scale and $M_i$, $m_i$ are soft supersymmetry
breaking mass parameters of order $\MS$, as 
defined in Ref.~\cite{Martin:1993zk}. 
In order for the left-hand side of Eq.~\eqref{mucond} to obtain the experimental
value, the first and the last term 
must cancel to a very large degree. There is no fundamental reason
why this is the case and the terms must be tuned. 

In practice, \HSSUSY\ 
inverts the Higgs minimisation equations, taking $\mu(\MS)$ and the value of
the \CP-odd Higgs boson mass $m_A(\MS)$ as input values. 
In this scheme, $m_{\bar H_i}^2$ are implicitly tuned in order to give $M_Z^2$
at 
the experimental central value. Once this single tuning has been achieved,
there are no more large corrections to $M_{h}$ from heavy sparticles: they
are all proportional to $M_Z \propto v$, which has
already been tuned.

\begin{figure}
  \includegraphics[width=\columnwidth]{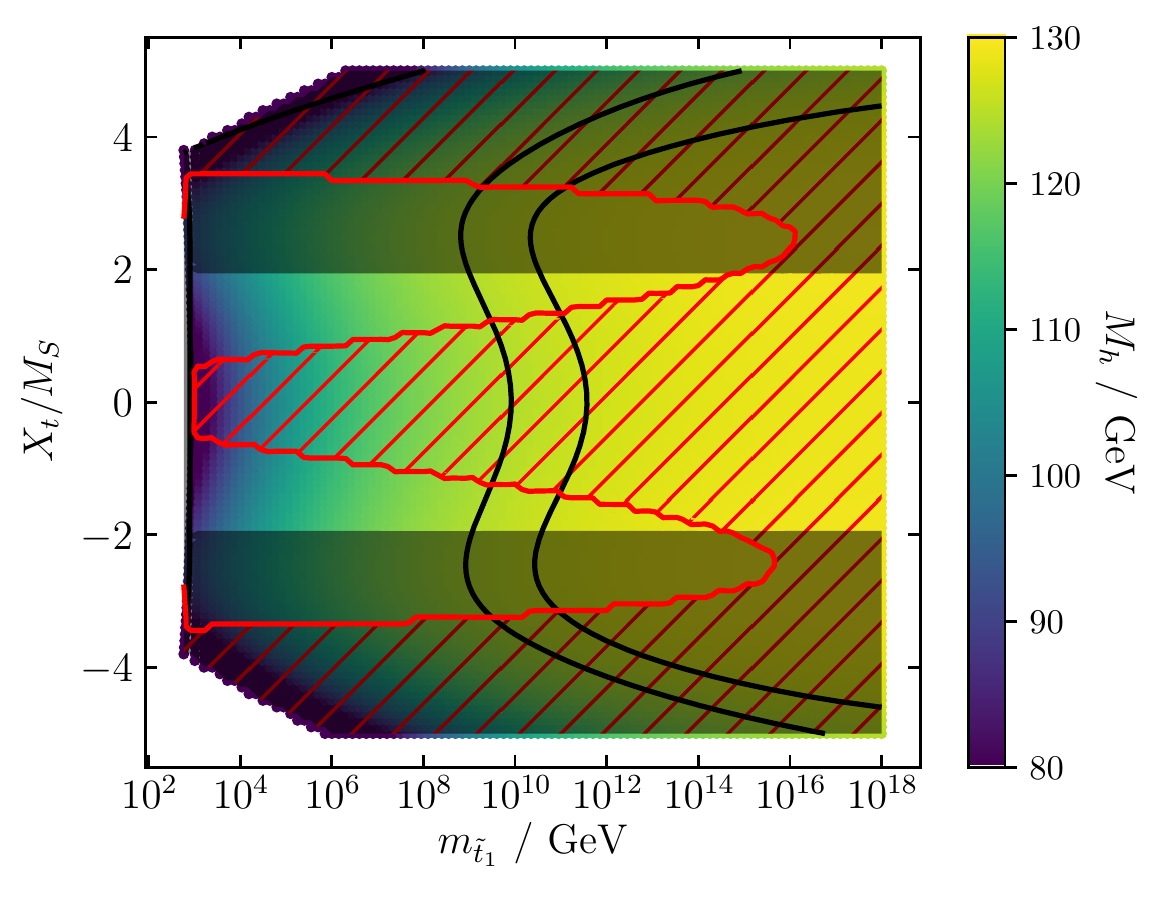}
  \caption{Higgs boson mass prediction with the the EFT calculation \HSSUSY\
    as a function of the lightest \DRbarp\ stop mass and the \DRbarp\
    stop mixing parameter for degenerate SUSY mass parameters and
    $\tan\beta(\MS) = 1$. The solid lines show the central value of $M_{h}$
  according to Eq.~\protect\eqref{higgsMass} plus or minus the theoretical
  uncertainty. The dark regions at the top and bottom of the plot display
  regions of parameter space which have charge and colour breaking
  minima. Hatched regions have $\lambda(\MS) < 0$.} 
  \label{fig:Mh_mstop1_xt}
\end{figure}

We estimate the upper bound on the lightest \DRbarp\ stop mass
$m_{\tilde{t}_1}$ in \fig{fig:Mh_mstop1_xt} by scanning over
$\MS$ and the relative \DRbarp\ stop mixing parameter
$X_t/\MS$ in a scenario with degenerate SUSY breaking mass parameters (except
for $m_{H_i}^2$) set equal to $\MS$, $\mu(\MS)=m_A(\MS)=\MS$ and 
$\tan\beta = 1$ to make the tree-level Higgs mass vanish. This should then be
the limiting case where we require the largest correction from stops in order
to predict $M_h$ in the correct range to satisfy the experimental measurement.
The Higgs boson mass has been calculated with the pure EFT calculation
\HSSUSY, because it has a smaller uncertainty than the fixed-order
calculations in the limit of large stop masses.  The black
lines show the contours of $M_h = 125.09\GeV \pm \DMhHSSUSY$, where
$\DMhHSSUSY$ is the estimate of the theory uncertainty from \HSSUSY,
as described in \secref{sec:1}.  In the red hatched region the quartic
Higgs coupling is negative at the SUSY scale, $\lambda(\MS) < 0$,
pointing to a potentially unstable electroweak vacuum\footnote{Around
  $X_t \approx 0$ the Higgsinos and electroweak gauginos give the
  dominant negative contribution to $\lambda(\MS)$ for
  $\tan\beta = 1$.  For slightly larger values of $X_t$ the stop
  contributions become dominant, leading to a positive $\lambda(\MS)$.
  For large stop mixing, the stop contribution becomes negative as
  well, driving $\lambda(\MS) < 0$ again.}\footnote{For slightly
  larger values of $\tan\beta$ the region around $X_t \approx 0$
  becomes allowed.  However, with larger $\tan\beta$ the
  tree-level Higgs boson mass rapidly increases, which leads to a
  significantly lower bound on the lightest stop mass.}.
In order to tell whether a point with $\lambda(\MS)<0$ really is excluded, one
should examine 
the full MSSM scalar potential. We consider this to be beyond the scope of
the present work, and so for now, we simply leave it as a point of interest. 
We also display regions which are excluded because they would lead to
charge and colour breaking minima which are deeper than the
electroweak vacuum, outside the region~\cite{Bagnaschi:2014rsa}
\begin{equation}
\frac{X_t^2}{m_{\tilde{Q}_3} m_{\tilde{u}_3}} < \left(4 - \frac{1}{\sin^2\beta}\right) \left(\frac{m_{\tilde{Q}_3}}{m_{\tilde{u}_3}} + \frac{m_{\tilde{u}_3}}{m_{\tilde{Q}_3}}\right) .
\label{CCB}
\end{equation}
Applying Eq.~\eqref{CCB} at
$Q=\MS$ with our boundary conditions on the quantities within it leads to 
\begin{equation}
-2 < X_t/\MS < 2,
 \label{xtCond}
\end{equation}
which corresponds to the non-darkened region in the horizontal middle band of
Fig.~\ref{fig:2}.
From regions with a stable electroweak vacuum based on the criterion in
Eq.~\eqref{xtCond} and where
Eq.~\eqref{higgsMass} is satisfied including the theoretical
uncertainty, we estimate an upper bound of
$m_{\tilde{t}_1} < 3.7 \times 10^{11}\GeV$.

\section{Summary}
\label{sec:3}

By using the state-of-the-art EFT Higgs boson mass prediction of
\HSSUSY\ we derived an estimate for the upper bound of the lightest
running stop mass that is compatible with the measured value of the
Higgs boson mass of $M_h = 125.09$ GeV, taking into account the
uncertainty estimate of the EFT calculation.  Our estimate for the
range of the lightest stop mass is
\begin{equation}
  m_{{\tilde t}_1} < 3.7 \times 10^{11} \GeV,
\end{equation}
provided the sparticle spectrum is not split so that some sparticles are much
lighter than $m_{{\tilde t}_1}$, as this would invalidate the
assumptions implicit within the EFT calculation that all sparticles are around
$\MS$. Our more precise
estimate of theoretical uncertainties in the prediction of $M_h$ does not qualitatively
change the conclusion of the previous study in Ref.~\cite{Bagnaschi:2014rsa}.
Unfortunately, such a bound is very much 
higher than the potential energies of conceivable terrestrial particle
accelerators. 

We also compared the precision of the Higgs boson mass
predictions of the state-of-the-art \DRbarp\ fixed-order and EFT
spectrum generators \SOFTSUSY, \FLEXIBLESUSY\ and \HSSUSY\ in the
\MSSM.  We estimated the uncertainties
of the Higgs boson mass of the fixed-order and the EFT calculation by
considering unknown logarithmic and non-logarithmic higher-order
corrections. As part of our work, we have provided a scheme to
estimate the theoretical uncertainties in fixed-order \DRbarp\
calculations, based on the prescription used in
Ref.~\cite{Athron:2016fuq}.  Our prescription is an extension of
Ref.~\cite{Athron:2016fuq}, which takes further sources of uncertainty
into account.
By comparing the precision of the predictions of the two
methods, we concluded that for SUSY masses below $\MS^\text{equal}=1.2\TeV$, the
fixed-order calculation is more precise, while above that scale the
EFT method is more precise. 
To estimate this scale, we took the  maximal mixing case where all soft
supersymmetry breaking 
masses are set to be degenerate at $\MS$ (except for $m_{H_i}$, which are
fixed in order to achieve successful EWSB) and where $\tan \beta=20$. The
precise value of $\MS^\text{equal}$ will change depending upon the scenario
and can vary between $\MS = 1.0\TeV$ and $1.3\TeV$ for minimal/maximal stop
mixing and small/large values of $\tan\beta$. 
However, once one imposes the experimental measurement upon $M_h$, $\MS \geq
1.3\TeV$ according to the fixed-order calculation\footnote{For a positive
  $X_t$ and varying $\tan \beta$, this may be reduced slightly to 1.1$\TeV$.}
and $2 \TeV$ according to 
the EFT calculation, as Fig.~\ref{fig:1} shows. For $\MS \geq 1.3\TeV$, the EFT has smaller
uncertainties 
and so one is likely to be in a 
r\'{e}gime where $M_h$ is better approximated by EFT methods. It is unclear as
yet, however, whether details of the \MSSM\ spectrum other than $M_h$ are better
approximated by EFT methods. 
One question which we have not addressed is: which approximation scheme
(fixed order \DRbarp\ or EFT) is more accurate
when there is a hierarchical sparticle spectrum? It is quite possible, for
example, that the stops are heavy but several of the other \MSSM\ sparticles are
significantly lighter.
For such scenarios the precision of the fixed order \DRbarp\ calculation would
have to be compared with the precision of an appropriate EFT that
contains the light sparticles.  We leave such a study for future work.

\begin{acknowledgements}
  We would like to thank the KUTS series of workshops, and LPTHE Paris
  for hospitality extended during the commencement of this work.
  A.V.\ would like to thank Emanuele Bagnaschi, Pietro Slavich and
  Georg Weiglein for many helpful discussions on the Higgs boson mass
  uncertainty estimate and Emanuele Bagnaschi for providing the two-
  and three-loop shifts to re-pa\-ra\-me\-tri\-se the quartic Higgs
  coupling in terms of the MSSM top Yukawa coupling. We thank Pietro Slavich
  for detailed comments on the manuscript.
  B.C.A.\ would like to
  thank the Cambridge SUSY Working Group for helpful discussions.
\end{acknowledgements}

\bibliographystyle{spphys}       
\bibliography{higgs.bib}   

\begin{thebibliography}{10}
\providecommand{\url}[1]{{#1}}
\providecommand{\urlprefix}{URL }
\expandafter\ifx\csname urlstyle\endcsname\relax
  \providecommand{\doi}[1]{DOI \discretionary{}{}{}#1}\else
  \providecommand{\doi}{DOI \discretionary{}{}{}\begingroup
  \urlstyle{rm}\Url}\fi

\bibitem{Aad:2012tfa}
G.~Aad, et~al., Phys. Lett. \textbf{B716}, 1 (2012).
\newblock \doi{10.1016/j.physletb.2012.08.020}

\bibitem{Chatrchyan:2012xdj}
S.~Chatrchyan, et~al., Phys. Lett. \textbf{B716}, 30 (2012).
\newblock \doi{10.1016/j.physletb.2012.08.021}

\bibitem{Aad:2015zhl}
G.~Aad, et~al., Phys. Rev. Lett. \textbf{114}, 191803 (2015).
\newblock \doi{10.1103/PhysRevLett.114.191803}

\bibitem{Allanach:2004rh}
B.C. Allanach, A.~Djouadi, J.L. Kneur, W.~Porod, P.~Slavich, JHEP \textbf{09},
  044 (2004).
\newblock \doi{10.1088/1126-6708/2004/09/044}

\bibitem{Siegel:1979wq}
W.~Siegel, Phys. Lett. \textbf{84B}, 193 (1979).
\newblock \doi{10.1016/0370-2693(79)90282-X}

\bibitem{Capper:1979ns}
D.M. Capper, D.R.T. Jones, P.~van Nieuwenhuizen, Nucl. Phys. \textbf{B167}, 479
  (1980).
\newblock \doi{10.1016/0550-3213(80)90244-8}

\bibitem{Jack:1994rk}
I.~Jack, D.R.T. Jones, S.P. Martin, M.T. Vaughn, Y.~Yamada, Phys. Rev.
  \textbf{D50}, R5481 (1994).
\newblock \doi{10.1103/PhysRevD.50.R5481}

\bibitem{Hempfling:1993qq}
R.~Hempfling, A.H. Hoang, Phys. Lett. \textbf{B331}, 99 (1994).
\newblock \doi{10.1016/0370-2693(94)90948-2}

\bibitem{Heinemeyer:1998kz}
S.~Heinemeyer, W.~Hollik, G.~Weiglein, Phys. Lett. \textbf{B440}, 296 (1998).
\newblock \doi{10.1016/S0370-2693(98)01116-2}

\bibitem{Heinemeyer:1998jw}
S.~Heinemeyer, W.~Hollik, G.~Weiglein, Phys. Rev. \textbf{D58}, 091701 (1998).
\newblock \doi{10.1103/PhysRevD.58.091701}

\bibitem{Heinemeyer:1998np}
S.~Heinemeyer, W.~Hollik, G.~Weiglein, Eur. Phys. J. \textbf{C9}, 343 (1999).
\newblock \doi{10.1007/s100529900006}

\bibitem{Heinemeyer:1999be}
S.~Heinemeyer, W.~Hollik, G.~Weiglein, Phys. Lett. \textbf{B455}, 179 (1999).
\newblock \doi{10.1016/S0370-2693(99)00417-7}

\bibitem{Degrassi:2001yf}
G.~Degrassi, P.~Slavich, F.~Zwirner, Nucl. Phys. \textbf{B611}, 403 (2001).
\newblock \doi{10.1016/S0550-3213(01)00343-1}

\bibitem{Brignole:2001jy}
A.~Brignole, G.~Degrassi, P.~Slavich, F.~Zwirner, Nucl. Phys. \textbf{B631},
  195 (2002).
\newblock \doi{10.1016/S0550-3213(02)00184-0}

\bibitem{Dedes:2003km}
A.~Dedes, G.~Degrassi, P.~Slavich, Nucl. Phys. \textbf{B672}, 144 (2003).
\newblock \doi{10.1016/j.nuclphysb.2003.08.033}

\bibitem{Heinemeyer:2004xw}
S.~Heinemeyer, W.~Hollik, H.~Rzehak, G.~Weiglein, Eur. Phys. J. \textbf{C39},
  465 (2005).
\newblock \doi{10.1140/epjc/s2005-02112-6}

\bibitem{Heinemeyer:2007aq}
S.~Heinemeyer, W.~Hollik, H.~Rzehak, G.~Weiglein, Phys. Lett. \textbf{B652},
  300 (2007).
\newblock \doi{10.1016/j.physletb.2007.07.030}

\bibitem{Hollik:2014bua}
W.~Hollik, S.~Pa{\ss}ehr, JHEP \textbf{10}, 171 (2014).
\newblock \doi{10.1007/JHEP10(2014)171}

\bibitem{Passehr:2017ufr}
S.~Pa{\ss}ehr, G.~Weiglein, Eur. Phys. J. \textbf{C78}(3), 222 (2018).
\newblock \doi{10.1140/epjc/s10052-018-5665-8}

\bibitem{Martin:2001vx}
S.P. Martin, Phys. Rev. \textbf{D65}, 116003 (2002).
\newblock \doi{10.1103/PhysRevD.65.116003}

\bibitem{Martin:2002iu}
S.P. Martin, Phys. Rev. \textbf{D66}, 096001 (2002).
\newblock \doi{10.1103/PhysRevD.66.096001}

\bibitem{Martin:2002wn}
S.P. Martin, Phys. Rev. \textbf{D67}, 095012 (2003).
\newblock \doi{10.1103/PhysRevD.67.095012}

\bibitem{Dedes:2002dy}
A.~Dedes, P.~Slavich, Nucl. Phys. \textbf{B657}, 333 (2003).
\newblock \doi{10.1016/S0550-3213(03)00173-1}

\bibitem{Brignole:2002bz}
A.~Brignole, G.~Degrassi, P.~Slavich, F.~Zwirner, Nucl. Phys. \textbf{B643}, 79
  (2002).
\newblock \doi{10.1016/S0550-3213(02)00748-4}

\bibitem{Martin:2003it}
S.P. Martin, Phys. Rev. \textbf{D70}, 016005 (2004).
\newblock \doi{10.1103/PhysRevD.70.016005}

\bibitem{Martin:2004kr}
S.P. Martin, Phys. Rev. \textbf{D71}, 016012 (2005).
\newblock \doi{10.1103/PhysRevD.71.016012}

\bibitem{Martin:2005eg}
S.P. Martin, Phys. Rev. \textbf{D71}, 116004 (2005).
\newblock \doi{10.1103/PhysRevD.71.116004}

\bibitem{Martin:2007pg}
S.P. Martin, Phys. Rev. \textbf{D75}, 055005 (2007).
\newblock \doi{10.1103/PhysRevD.75.055005}

\bibitem{Harlander:2008ju}
R.V. Harlander, P.~Kant, L.~Mihaila, M.~Steinhauser, Phys. Rev. Lett.
  \textbf{100}, 191602 (2008).
\newblock \doi{10.1103/PhysRevLett.101.039901, 10.1103/PhysRevLett.100.191602}.
\newblock [Phys. Rev. Lett.101,039901(2008)]

\bibitem{Kant:2010tf}
P.~Kant, R.V. Harlander, L.~Mihaila, M.~Steinhauser, JHEP \textbf{08}, 104
  (2010).
\newblock \doi{10.1007/JHEP08(2010)104}

\bibitem{Martin:2017lqn}
S.P. Martin, Phys. Rev. \textbf{D96}(9), 096005 (2017).
\newblock \doi{10.1103/PhysRevD.96.096005}

\bibitem{ATLAS:2017kyf}
{The ATLAS collaboration},  ATLAS-CONF-2017-020 (2017)

\bibitem{ATLAS:2017tpg}
{The ATLAS collaboration},  ATLAS-CONF-2017-019 (2017)

\bibitem{Gomez-Ceballos:2013zzn}
M.~Bicer, et~al., JHEP \textbf{01}, 164 (2014).
\newblock \doi{10.1007/JHEP01(2014)164}

\bibitem{Dam:2015nir}
M.~Dam, PoS \textbf{EPS-HEP2015}, 334 (2015)

\bibitem{Janot:2015mqv}
P.~Janot, PoS \textbf{EPS-HEP2015}, 333 (2015)

\bibitem{Giudice:2011cg}
G.F. Giudice, A.~Strumia, Nucl. Phys. \textbf{B858}, 63 (2012).
\newblock \doi{10.1016/j.nuclphysb.2012.01.001}

\bibitem{Draper:2013oza}
P.~Draper, G.~Lee, C.E.M. Wagner, Phys. Rev. \textbf{D89}(5), 055023 (2014).
\newblock \doi{10.1103/PhysRevD.89.055023}

\bibitem{Bagnaschi:2014rsa}
E.~Bagnaschi, G.F. Giudice, P.~Slavich, A.~Strumia, JHEP \textbf{09}, 092
  (2014).
\newblock \doi{10.1007/JHEP09(2014)092}

\bibitem{Lee:2015uza}
G.~Lee, C.E.M. Wagner, Phys. Rev. \textbf{D92}(7), 075032 (2015).
\newblock \doi{10.1103/PhysRevD.92.075032}

\bibitem{Vega:2015fna}
J.~Pardo~Vega, G.~Villadoro, JHEP \textbf{07}, 159 (2015).
\newblock \doi{10.1007/JHEP07(2015)159}

\bibitem{Allanach:2001kg}
B.C. Allanach, Comput. Phys. Commun. \textbf{143}, 305 (2002).
\newblock \doi{10.1016/S0010-4655(01)00460-X}

\bibitem{Allanach:2014nba}
B.C. Allanach, A.~Bednyakov, R.~Ruiz~de Austri, Comput. Phys. Commun.
  \textbf{189}, 192 (2015).
\newblock \doi{10.1016/j.cpc.2014.12.006}

\bibitem{Athron:2014yba}
P.~Athron, J.h. Park, D.~St{\"{o}}ckinger, A.~Voigt, Comput. Phys. Commun.
  \textbf{190}, 139 (2015).
\newblock \doi{10.1016/j.cpc.2014.12.020}

\bibitem{Athron:2017fvs}
P.~Athron, M.~Bach, D.~Harries, T.~Kwasnitza, J.h. Park, D.~St{\"{o}}ckinger,
  A.~Voigt, J.~Ziebell,   (2017)

\bibitem{Harlander:2017kuc}
R.V. Harlander, J.~Klappert, A.~Voigt, Eur. Phys. J. \textbf{C77}(12), 814
  (2017).
\newblock \doi{10.1140/epjc/s10052-017-5368-6}

\bibitem{Bahl:2017aev}
H.~Bahl, S.~Heinemeyer, W.~Hollik, G.~Weiglein, Eur. Phys. J. \textbf{C78}(1),
  57 (2018).
\newblock \doi{10.1140/epjc/s10052-018-5544-3}

\bibitem{Hahn:2013ria}
T.~Hahn, S.~Heinemeyer, W.~Hollik, H.~Rzehak, G.~Weiglein, Phys. Rev. Lett.
  \textbf{112}(14), 141801 (2014).
\newblock \doi{10.1103/PhysRevLett.112.141801}

\bibitem{Bahl:2016brp}
H.~Bahl, W.~Hollik, Eur. Phys. J. \textbf{C76}(9), 499 (2016).
\newblock \doi{10.1140/epjc/s10052-016-4354-8}

\bibitem{Athron:2016fuq}
P.~Athron, J.h. Park, T.~Steudtner, D.~St{\"{o}}ckinger, A.~Voigt, JHEP
  \textbf{01}, 079 (2017).
\newblock \doi{10.1007/JHEP01(2017)079}

\bibitem{Avdeev:1997sz}
L.V. Avdeev, M.{\relax Yu}. Kalmykov, Nucl. Phys. \textbf{B502}, 419 (1997).
\newblock \doi{10.1016/S0550-3213(97)00404-5}

\bibitem{Bednyakov:2002sf}
A.~Bednyakov, A.~Onishchenko, V.~Velizhanin, O.~Veretin, Eur. Phys. J.
  \textbf{C29}, 87 (2003).
\newblock \doi{10.1140/epjc/s2003-01178-4}

\bibitem{Bednyakov:2005kt}
A.~Bednyakov, D.I. Kazakov, A.~Sheplyakov, Phys. Atom. Nucl. \textbf{71}, 343
  (2008).
\newblock \doi{10.1007/s11450-008-2015-6}

\bibitem{Degrassi:2012ry}
G.~Degrassi, S.~Di~Vita, J.~Elias-Miro, J.R. Espinosa, G.F. Giudice,
  G.~Isidori, A.~Strumia, JHEP \textbf{08}, 098 (2012).
\newblock \doi{10.1007/JHEP08(2012)098}

\bibitem{Martin:2014cxa}
S.P. Martin, D.G. Robertson, Phys. Rev. \textbf{D90}(7), 073010 (2014).
\newblock \doi{10.1103/PhysRevD.90.073010}

\bibitem{Bagnaschi:2017xid}
E.~Bagnaschi, J.~Pardo~Vega, P.~Slavich, Eur. Phys. J. \textbf{C77}(5), 334
  (2017).
\newblock \doi{10.1140/epjc/s10052-017-4885-7}

\bibitem{Bednyakov:2013eba}
A.V. Bednyakov, A.F. Pikelner, V.N. Velizhanin, Nucl. Phys. \textbf{B875}, 552
  (2013).
\newblock \doi{10.1016/j.nuclphysb.2013.07.015}

\bibitem{Buttazzo:2013uya}
D.~Buttazzo, G.~Degrassi, P.P. Giardino, G.F. Giudice, F.~Sala, A.~Salvio,
  A.~Strumia, JHEP \textbf{12}, 089 (2013).
\newblock \doi{10.1007/JHEP12(2013)089}

\bibitem{Chetyrkin:1999qi}
K.G. Chetyrkin, M.~Steinhauser, Nucl. Phys. \textbf{B573}, 617 (2000).
\newblock \doi{10.1016/S0550-3213(99)00784-1}

\bibitem{Melnikov:2000qh}
K.~Melnikov, T.v. Ritbergen, Phys. Lett. \textbf{B482}, 99 (2000).
\newblock \doi{10.1016/S0370-2693(00)00507-4}

\bibitem{Chetyrkin:2000yt}
K.G. Chetyrkin, J.H. Kuhn, M.~Steinhauser, Comput. Phys. Commun. \textbf{133},
  43 (2000).
\newblock \doi{10.1016/S0010-4655(00)00155-7}

\bibitem{Martin:2015eia}
S.P. Martin, Phys. Rev. \textbf{D92}(5), 054029 (2015).
\newblock \doi{10.1103/PhysRevD.92.054029}

\bibitem{Chetyrkin:2016ruf}
K.G. Chetyrkin, M.F. Zoller, JHEP \textbf{06}, 175 (2016).
\newblock \doi{10.1007/JHEP06(2016)175}

\bibitem{Bednyakov:2015ooa}
A.V. Bednyakov, A.F. Pikelner, Phys. Lett. \textbf{B762}, 151 (2016).
\newblock \doi{10.1016/j.physletb.2016.09.007}

\bibitem{BVW}
E.~Bagnaschi, A.~Voigt, G.~Weiglein, {Higgs mass computation for hierarchical
  spectra in the MSSM using FlexibleSUSY}.
\newblock In preparation

\bibitem{Braathen:2017jvs}
J.~Braathen, M.D. Goodsell, M.E. Krauss, T.~Opferkuch, F.~Staub, Phys. Rev.
  \textbf{D97}(1), 015011 (2018).
\newblock \doi{10.1103/PhysRevD.97.015011}

\bibitem{Staub:2017jnp}
F.~Staub, W.~Porod, Eur. Phys. J. \textbf{C77}(5), 338 (2017).
\newblock \doi{10.1140/epjc/s10052-017-4893-7}

\bibitem{Bahl:2018jom}
H.~Bahl, W.~Hollik, arXiv:1805.00867  (2018)

\bibitem{Pierce:1996zz}
D.M. Pierce, J.A. Bagger, K.T. Matchev, R.j. Zhang, Nucl. Phys. \textbf{B491},
  3 (1997).
\newblock \doi{10.1016/S0550-3213(96)00683-9}

\bibitem{Martin:1993zk}
S.P. Martin, M.T. Vaughn, Phys. Rev. \textbf{D50}, 2282 (1994).
\newblock \doi{10.1103/PhysRevD.50.2282, 10.1103/PhysRevD.78.039903}.
\newblock [Erratum: Phys. Rev.D78,039903(2008)]

\end{thebibliography}

\end{document}